\documentclass[11pt,a4paper]{article}
\usepackage{jcappub}
\usepackage{graphicx}  
\usepackage{dcolumn}   
\usepackage{bm}        
\usepackage{caption}
\usepackage{amssymb}   
\usepackage{epsfig}
\usepackage{amsmath}
\usepackage{appendix}

\newcommand{\be}{\begin{equation}}
\newcommand{\ee}{\end{equation}}
\newcommand{\beq}{\begin{eqnarray}}
\newcommand{\eeq}{\end{eqnarray}}

\title{The generalized Chaplygin--Jacobi gas}
\author{ J. R. Villanueva\note{Corresponding author.} }
 \affiliation{\it Instituto de F\'{\i}sica y Astronom\'{\i}a, Universidad de
Valpara\'{\i}so, Gran Breta\~na  1111, Playa Ancha, Valpara\'{\i}so, Chile.}
\affiliation{\it Centro de Astrof\'isica de Valpara\'iso, Gran Breta\~na  1111, Playa Ancha, Valpara\'{\i}so, Chile.}
 \emailAdd{jose.villanuevalob@uv.cl}

\date{\today}

\abstract{The present paper is devoted to find a
new generalization of the generalized Chaplygin gas.
Therefore, starting from the Hubble parameter 
associated to the Chaplygin scalar field and
using some elliptic identities,
the elliptic generalization is straightforward. 
Thus, all relevant quantities that drive inflation
are calculated exactly. Finally, using
the measurement on inflation from the 
{\it Planck}  2015 results, observational constraints
on the parameters are given.}

\keywords{physics of the early universe, inflation, exact solutions.}

\begin{document}

\maketitle


\section{Introduction}\label{intro}
Clearly, evidence from the most varied 
observations suggest that near 25\% of
the cosmic energy--matter corresponds to the so-called
cold dark matter (CDM) and the remaining 75\%
corresponds to the so-called dark energy (DE).
The dark matter problem born with the Zwicky's
observation of a large velocity dispersion
of the components in the Coma cluster \cite{zwicky}
together with Babcock observation of galactic
rotation curves in the Andromeda galaxy \cite{babcock}.
This paradigm has been studied in several
context, for example from the particle physics point of view,
the candidates are
 axions, inert Higgs doublet, sterile neutrinos,
super--symmetric particles and Kaluza--Klein particles
\cite{jungman,bergstrom00,bertone05,bergstrom09,bertone09},
while from the geometrical point of view
Weyl's theory has received much attention 
from theoretical and experimental physicists
\cite{mannhein89,kazanas91,mannhein91,mannhein94,diaferio11,pireux1,pireux2,ov13}.
On the other hand, one of the most 
natural candidates to explain DE 
corresponds to a cosmological constant $\Lambda$, 
which, although successful as a fit model,
 it shows a huge discrepancy between 
its theoretical and observed values \cite{carroll}.
This evidence has promoted the study of new types of 
candidates that can link the different stage 
of the universe. One type of cosmic fluid is often 
called {\it quintessence} and has been widely used because
it behaves like a cosmological constant by combining positive 
energy density and negative pressure. Nevertheless, 
unlike the cosmological
constant, the pressure of 
this fluid is dynamic.
An example of this type is the so-called {\it Chaplygin gas} (CG)
which is an exotic fluid having the equation of state (EoS)
\begin{equation}
\label{chapn}
p_{ch}=-\frac{B}{\rho_{ch}},
\end{equation}
where $B$ is a positive constant, $p_{ch}$
and $\rho_{ch}$ are the pressure and energy density
in a comoving reference frame, respectively.
This EoS was originally proposed
to describe the lifting force on a wing of an airplane 
in aerodynamics \cite{chaplygin}. 
In cosmology, it was first introduce by  \cite{Kamenshchik:2001cp,Bilic:2001cg,Fabris:2001tm},
nevertheless, 
in its original form, this fluid 
presents inconsistency with 
some observational data \cite{makler,zhu,sen} 
such that a generalization is necessary to improve 
consistency with data. A first attempt was made considering the 
{\it generalized Chaplygin gas} (GCG) whose EoS 
is given by \cite{bilic,bento02,gorini,bento04}
\begin{equation}
\label{ede}
p_{gcg}=-\frac{B}{\rho_{gcg}^{\alpha}},
\end{equation}
where $B>0$ and the GCG parameter
lies in the range $0\leq\alpha\leq 1$. Obviously,
the CG is recovered by making $\alpha=1$, while the case
$\alpha=0$ mimics the effect of a cosmological constant.
An important characteristic of the 
Chaplygin gas (CG and GCG) 
is that this interpolates
between a dust dominated 
phase where $\rho\propto a^{-3}$,
and a de Sitter phase where $p=-\rho$.
The different parameters that enter into the model 
has been confronted successfully with many observational data 
(see, for example \cite{dcv,liang,freitas}),
but, unfortunately, the model 
produce oscillations or exponential blowup
in the matter power spectrum, 
being inconsistent with observation \cite{sandvik}.

On the other hand, an approach to single-field 
inflation using the GCG as inflaton field was performed 
by del Campo \cite{delcampo13}.
By comparing with the measurement 
for the scalar spectral index together
with its running
from the
{\it Planck} 2013 data \cite{Planck:2013jfk}, 
 del Campo found that 
the best value for the $\alpha$--parameter 
is given by $\alpha=0.2578\pm0.0009$.
Therefore, following the idea outlined in \cite{VG15}, 
the main goal of this paper 
is to give a new generalization
for the Chaplygin scalar field 
by using elliptic functions to describe
the inflationary stage and then
perform the confrontation between the model and 
the measurement recently released by the Planck 2015 data.

The outline of this work is as follows:
the basic aspects to describe exact solutions 
to inflationary universe models within the 
framework of the Hamilton--Jacobi approach to cosmology,
together with review of the results found earlier by del Campo
for the GCG are showed in Section II.
In Section III the Chaplygin--Jacobi gas is presented and 
the Hamilton--Jacobi formalism is applied to
find the most relevant quantities in the inflationary stage.
Finally, in Section IV conclusions and final remarks are presented.
In addition, Appendix A few useful formulas 
and identities that satisfy the Jacobi elliptic functions 
are shown.

\section{The generalized Chaplygin gas in inflation}
The fundamental quantity in inflation is the scalar 
{\it inflaton} field $\phi$ which leads to the universe 
to expand extremely rapid  in a very short time.
The evolution of this field becomes governed 
by its scalar potential, $V(\phi)$, 
via the Klein--Gordon equation
\begin{equation}
\label{kg}
\ddot{\phi}+3H\dot{\phi}+V_{,\phi}=0,
\end{equation} 
where $V_{,\phi}$ represents a derivative with respect to $\phi$.
Thus, this equation of motion, together with the Friedmann 
equation, 
\begin{equation}
\label{fried}
H^2=\frac{8\pi}{3 m_p^2}\left[\frac{1}{2}\dot{\phi}^2+V(\phi)\right]
\end{equation}
obtained from Einstein general relativity theory, 
form the most simple set of field equations, 
which could be applied to obtain inflationary solutions.
The set of equations (\ref{kg}-\ref{fried}) are the mainstay
of Hamilton--Jacobi approach 
\cite{carr93,Schunck,hawkins01,Kinney:1997ne,delCampo:2012qb,Kim,chaadaev}.

In order to obtain exact solutions in this approach,
del Campo \cite{delcampo13}  and Dinda et al \cite{dinda14}
used the Chaplygin gas whose generating 
function is given by
\begin{equation}
H(\phi)=H_0 \cosh^{\frac{1}{1+\alpha}}\left[(1+\alpha)\Phi)\right],
\label{gener}
\end{equation}
where $\Phi\equiv \sqrt{\frac{6\pi}{m_p^2}}
(\phi-\phi_0)$ is a dimensionless scalar field and
$H_0$, the value of $H$ when $\phi=\phi_0$,
is a constant given by
$H_0=\sqrt{\frac{8\pi}{3 m_p^2}}\,B^{\frac{1}{2(1+\alpha)}}$.
From the generating function (\ref{gener}) 
one can obtain all the relevant quantities for inflation,
for example the scale factor
\begin{equation}
\label{b8}
a(\phi)=a_i\,\exp\left[-\frac{4\pi}{m_{p}^2}\,\int_{\phi_i}^{\phi}
\frac{H(\phi)}{H_{,\phi}(\phi)} d\phi\right],
\end{equation}
which becomes \footnote{the original paper \cite{delcampo13}
presents a typo in this formulae.}
\begin{equation}
\label{fde}
a(\phi)=a_i\,\left(\frac{\sinh[(1+\alpha)\Phi]}{\sinh[(1+\alpha)\Phi_i]}\right)^{\frac{2}{3(1+\alpha)}}.
\end{equation}
The number of e-folding of physical expansion that occur 
in the inflationary stage is given by
\begin{equation}
\label{b12}
N\equiv \int_{t_{i}}^{t_f} H \,dt=\frac{4\pi}{m_p^2}
\int_{\phi_f}^{\phi_i}\frac{H}{H_{,\phi}}\,d\phi,
\end{equation}
so, del Campo found that for the  GCG 
this quantity becomes\footnote{the original paper \cite{delcampo13}
presents a typo in this formulae.}
\begin{equation}
\label{efoldgcg}
N(\phi)=\frac{2}{3(1+\alpha)}\,\mathrm{ln}\left \{\frac{\sinh[(1+\alpha)\Phi_e]}{\sinh[(1+\alpha)\Phi]}
\right\},
\end{equation}
while the scalar potential 
\begin{equation}
\label{potHJ}
V(\phi)=\left(\frac{3 m_p^2}{8\pi}\right)
\left[H^2
-\frac{m_p^2}{12\pi}H_{,\phi}^2\right],
\end{equation}
is given by the following expression:
\begin{equation}
\label{potgcg}
V(\phi)=V_0\, \frac{1+\cosh^2[(1+\alpha) \Phi]}{\cosh^{\frac{2\alpha}{1+\alpha}}[(1+\alpha)\Phi]},
\end{equation}
where $V_0=\frac{1}{2}B^{\frac{1}{1+\alpha}}$. 
In the slow--rolls approximation this scalar 
potential is well behaved and reduces to
\begin{equation}
\label{spsr}
V_{s-r}(\phi)\simeq V_0 \cosh^{\frac{2}{1+\alpha}}[(1+\alpha)\Phi].
\end{equation}

In next section, the Hamilton--Jacobi formalism 
is applied to obtain exact solutions
in the inflationary stage using a elliptic 
generalization of the GCG.

\section{The generalized Chaplygin--Jacobi gas}

It is well known that trigonometric and 
hyperbolic functions are special cases of 
more general functions, the so-called 
elliptic functions, which are doubly periodic 
in the complex plane, while the above 
are periodic on the real axis and the 
imaginary axis, respectively.
With this 
in mind and using the rules 
given in appendix \ref{app:jef},
the generating function (\ref{gener}) is
conveniently written as
\begin{equation}
\label{p2.3}
H(\phi, k)=H_0\,\textrm{nc}^{\frac{1}{1+\alpha}}\left[(1+\alpha)\,\Phi\right],
\end{equation}
where $\textrm{nc}(x)=1/\textrm{cn}(x)$, 
and $\textrm{cn}(x)\equiv\textrm{cn}(x|\, k)$ 
is the Jacobi elliptic cosine function, 
and $k$ is the modulus. 
This choice allows to obtain the 
generating function of the GCG (\ref{gener})
making $k\rightarrow 1$ in equation (\ref{p2.3}).
Henceforth we denote this scalar field as the
{\it generalized Chaplygin--Jacobi gas} (GCJG).

Taking into account that pressure and energy density are given by
\begin{equation}
\label{dens1}
p_{\phi}=\frac{3 m_p^2}{8\pi} H^2 \left[\frac{m_p^2}{6\pi}\left(\frac{H_{,\phi}}{H}\right)^2
-1\right]
\end{equation}
and
\begin{equation}
\label{press1}
\rho_{\phi}=\frac{3 m_p^2}{8\pi} H^2,
\end{equation}
respectively, and by using the generating function (\ref{p2.3})
it is no hard to obtain that
\begin{equation}
\label{press}
p_{\phi}=-\frac{B^\frac{1}{2(1+\alpha)}}{k}\,\text{nc}^{\frac{2(2+\alpha)}{1+\alpha}}
\left[(1+\alpha)\,\Phi\right]\,(\text{dn}^{4}
\left[(1+\alpha)\,\Phi\right]-k'),
\end{equation}
and
\begin{equation}
\label{dens}
\rho_{\phi}=B^\frac{1}{2(1+\alpha)}\,\textrm{nc}^{\frac{2}{1+\alpha}}\left[(1+\alpha)\,\Phi\right],
\end{equation}
where $k'=1-k$ is the complementary modulus.
Therefore, by using eq. (\ref{dens}) one obtain
$\textrm{nc}\left[(1+\alpha)\,\Phi\right]$ as a function
of $\rho_{\phi}$ and then, by substituting
into eq. (\ref{press}) together with some identities
of the Jacobi elliptic functions,
the EoS related to the GCJG read
\begin{equation}\label{prro}
p_{\phi}=-\frac{B\,k}{\rho_{\phi}^{\alpha}}
-2\,k'\,\rho_{\phi}+\frac{k'}{B}\,\rho_{\phi}^{2+\alpha}
\end{equation}
Obviously,  in the limit 
$k\rightarrow 1$ (or $k'\rightarrow 0$) this EoS
coincide with (\ref{ede}).
\begin{figure}
  \begin{center}
    \includegraphics[width=150mm]{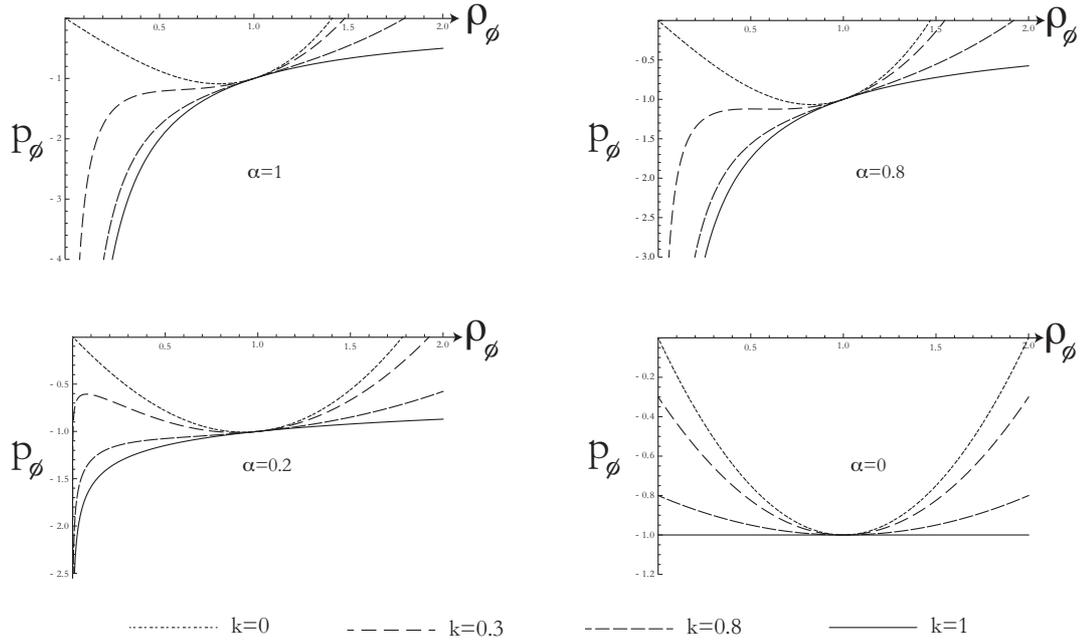}
  \end{center}
  \caption{Plot of the pressure against density
 for generalized Chaplygin--Jacobi gas (\ref{prro}). Here we
 plot different values of the modulus $k$ and GCG parameter $\alpha$.
 In all graphs was used $B=1$. }
    \label{eoscj}
\end{figure}
Note that, for $k\neq 1$, the GCJG presents a novel
behaviour in which the pressure is a positive quantity. 
This fact occurs
at the range $\rho_{\phi}^{(c)}<\rho_{\phi}$, where
critical density,
$\rho_{\phi}^{(c)}$, depends on the parameters across
the following expression
\begin{equation}
\label{rhocr}
\rho_{\phi}^{(c)}
=
\left[ B\left(1+\frac{1}{\sqrt{k'}}\right)\right]^{\frac{1}{1+\alpha}}.
\end{equation}
At this point $p_{\phi}=0$, and the fluid 
presents a dust--like behaviour.
An interesting feature of this EoS appears 
when $k=0$ (or $k'=1$) which corresponds to 
the trigonometric limit. In addition to the root 
obtained from equation (\ref{rhocr}), must 
also add $\rho_{\phi}^{(c)}=0$, and therefore
the GCJG presents a dust--like behaviour
when $\rho_{\phi}=0$ and $\rho_{\phi}=(2\,B)^{\frac{1}{1+\alpha}}$,
as shown in Figure \ref{eoscj}.

For this fluid one has simply that the EoS parameter becomes
\begin{equation}
\label{gcjgp}
\omega_{\phi}=\frac{p_{\phi}}{\rho_{\phi}}=
\frac{k'}{B}\rho_{\phi}^{1+\alpha}-
\frac{B\,k}{\rho_{\phi}^{1+\alpha}}-2\,k',
\end{equation}
while the square 
of the speed of sound is
\begin{equation}
\label{sos}
v_{\phi}^2=\frac{\partial p_{\phi}}{\partial \rho_{\phi}}=
\frac{(2+\alpha)\,k'}{B}\rho_{\phi}^{1+\alpha}+
\frac{B\,k\,\alpha}{\rho_{\phi}^{1+\alpha}}-2\,k'.
\end{equation}
Note that the speed of sound vanished 
when the density reaches the value
\begin{equation}
\label{soscero}
\rho_{\phi}^{(0)}
=
\left[ B\left(\frac{1\pm\sqrt{1-\frac{k}{k'}\alpha(2+\alpha)}}{2+\alpha}\right)\right]^{\frac{1}{1+\alpha}}.
\end{equation}
Of course, the existence 
of these points is restricted 
to pairs that satisfy the condition
$k^{-1}>(1+\alpha)^2$. For example, 
if $\alpha=0$, then $k<1$, while
for the case $\alpha=1$
the modulus must be $k<0.25$.
It is important to note that 
null speed of sound do not implies 
null pressure. In fact, at this point
the pressure is
\begin{equation}
\label{presssos1}
p_{\phi 0}\equiv p_{\phi}(\rho_{\phi}=\rho_{\phi}^{(0)})
=
\frac{2B(1+\alpha)}{(2+\alpha)^2\,(\rho_{\phi}^{(0)})^{\alpha}}
\left\{1+k(1+\alpha)\pm \sqrt{k'\,[1-k(1+\alpha)^2]} \right\}.
\end{equation}

\begin{figure}
  \begin{center}
    \includegraphics[width=150mm]{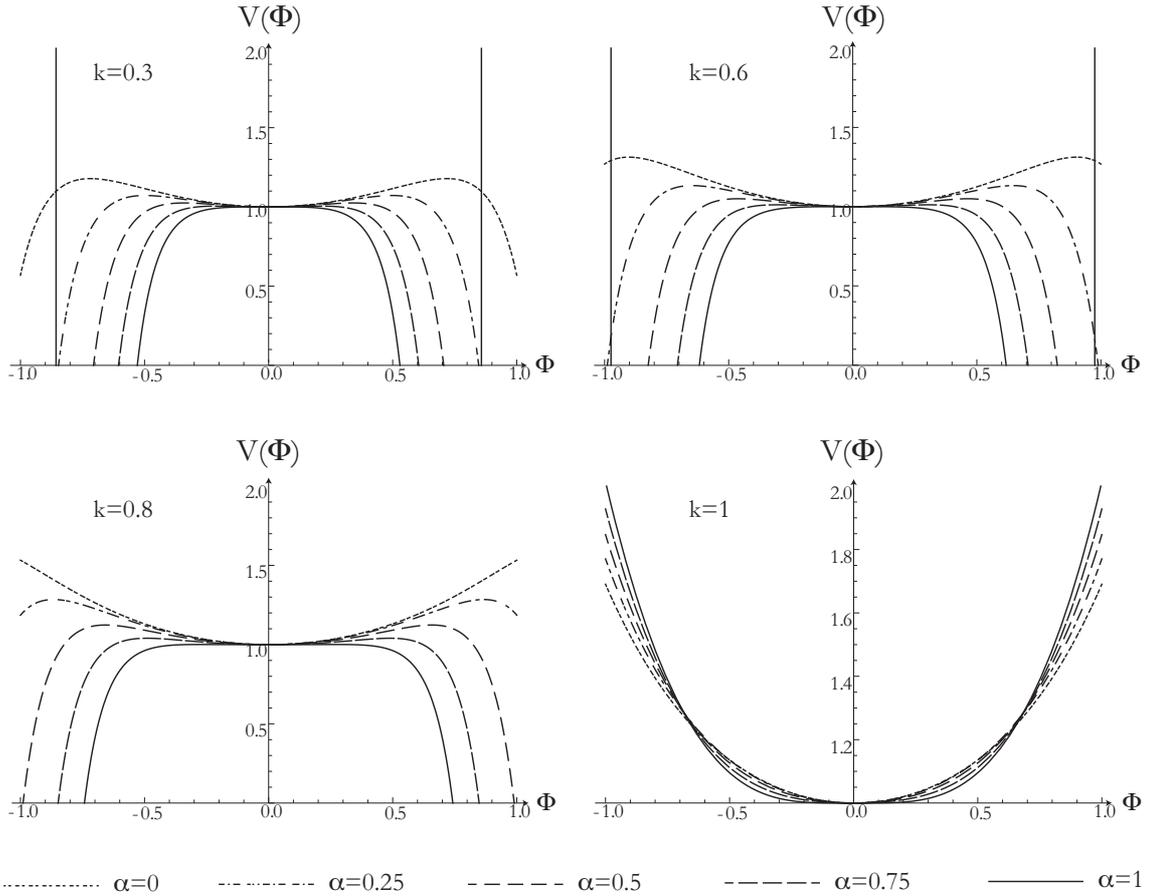}
  \end{center}
  \caption{Plots of the scalar potentials
 for generalized Chaplygin--Jacobi gas, $V(\Phi)$,
 as a function of the dimensionless scalar field, 
 $\Phi\equiv \sqrt{\frac{6\pi}{m_p^2}(\phi-\phi_0)}$. Here the
 scalar potential $V(\Phi)$ is expressed as a multiple
 of the constant $V_0\equiv \frac{1}{2}B^{\frac{1}{1+\alpha}}$.}
    \label{fpot}
\end{figure}

On the other hand, using eq. (\ref{potHJ}) together with eq. (\ref{p2.3})
one obtain for the scalar potential
\begin{equation}
\label{pot1}
V(\phi)=V_0 \frac{2\text{nc}^{2}\left[(1+\alpha)\,\Phi\right]-\text{dc}^{2}\left[(1+\alpha)\,\Phi\right]\,\text{sc}^{2}\left[(1+\alpha)\,\Phi\right]}{\textrm{nc}^{\frac{2\alpha}{1+\alpha}}\left[(1+\alpha)\,\Phi\right] },
\end{equation}
where $V_0=\frac{1}{2}B^\frac{1}{2(1+\alpha)}$. This scalar potential is depicted
in Figure \ref{fpot} for different values of the GCG parameter $\alpha$ and the modulus $k$.
Also, by using eqs. (\ref{potHJ})
and (\ref{p2.3}) is straightforward
to found the potential as a function 
of the generating function, so
\begin{equation}
\label{potvsH}
V(H)=V_0
\left(\frac{H}{H_0}\right)^{2}
\left\{
2-
\left[
k' \left(\frac{H}{H_0}\right)^{2(1+\alpha)}+k
\right]
\left[
1-\left(\frac{H_0}{H}\right)^{2(1+\alpha)}
\right]
\right\}.
\end{equation}

By combining the generating function (\ref{p2.3}) with  
eq. (\ref{b8}) allows to find the scale factor as
a function of the inflaton field:
\begin{equation}
\label{ainf}
a(\phi)=
a_i\,
\left(\frac{\textrm{sd}[(1+\alpha)\Phi]}{\textrm{sd}[(1+\alpha)\Phi_i]}\right)^{\frac{2}{3(1+\alpha)}}.
\end{equation}
Here $\textrm{sd}(x)\equiv \textrm{sd}(x|\,k)=\textrm{sn}(x)/\textrm{dn}(x)$,
where $\textrm{sn}(x)\equiv \textrm{sn}(x|\,k)$ is the Jacobi elliptic sine function
and $\textrm{dn}(x)\equiv \textrm{dn}(x|\,k)$ is the Jacobi elliptic delta function.

With the aim to verify that an inflationary period occurs, it 
is very helpful to introduce the deceleration parameter 
$q=\frac{m_p^2}{4\pi}\left(\frac{H_{,\phi}}{H}\right)^2-1$,
which becomes
\begin{equation}
\label{qcj}
q=\frac{3}{2}\frac{\text{dn}^{2}\left[(1+\alpha)\,\Phi\right]\,
\text{sn}^{2}\left[(1+\alpha)\,\Phi\right]}{\text{cn}^{2}\left[(1+\alpha)\,\Phi\right]}
-1.
\end{equation}
Making use of some elliptic identities 
it is possible obtain the value for $\phi_e$, the value for the 
scalar field when inflation ends (i. e. $q(\phi_e)=0$).
For example, if eq. (\ref{qcj}) is writes in terms of
the Jacobi elliptic sine one arrives to the following
quadratic equation
\begin{equation}
\label{equad}
3k\,y^2-5\,y+2=0,
\end{equation}
where $y=\text{sn}^{2}\left[(1+\alpha)\,\Phi_e\right]$.
Therefore, solving for $\Phi_e$ one get that
the four roots are given by
\begin{equation}
\label{phie}
\Phi_e=\frac{F\left[\arcsin \left(\pm \sqrt{ y_{\pm}}\right),\,k\right]}{1+\alpha},
\end{equation}
where $F(\varphi,\,k)$ is the normal elliptic integral of the first kind
and
\begin{equation}
\label{yk}
y_{\pm}=\frac{5}{6k}\left(1\pm \sqrt{1-\frac{24\,k}{25}}\right).
\end{equation}
It is important to note that real solutions are found
for all $0\leq k \leq 1$. Also, the condition
$y\leq 1$ implies that the allowed solution
are $\pm \sqrt{y_-}$.

On the other hand,  using eq. (\ref{b12}) one get that 
the number of the e-folds becomes
\begin{equation}
\label{efcj}
N(\phi)=\frac{2}{3(1+\alpha)}\ln\left\{\frac{\textrm{sd}[(1+\alpha)\Phi_e]}{\textrm{sd}[(1+\alpha)\Phi]}\right\},
\end{equation}
where $\Phi_e$ is given by eq. (\ref{phie}).
For example, for $N=70$ the above leads to 
\begin{equation}
\label{n70}
\Phi_{70}=\frac{F\left[\arcsin\left(\frac{x}{\sqrt{1+k x^2}}\right),\,k\right]}{1+\alpha},
\end{equation}
where $x$ is given by
\begin{equation}
\label{x70}
x=
\textrm{sd}[(1+\alpha)\Phi_e]\,e^{-105(1+\alpha)}.
\end{equation}

In order to obtain some information 
about the perturbations, 
result instructive to define
the three first hierarchy parameters
in terms of the generating function and its derivatives.
Thus,  the first Hubble hierarchy parameter  $\epsilon_H$ 
is given by
\begin{equation}\label{b10}
\epsilon_H\equiv - \frac{d \ln H}{d \ln a}=\left(\frac{m_p^2}{4\pi}\right)
\left(\frac{H_{,\phi}}{H}\right)^2.
\end{equation}
Note that
this parameter provides information 
about the acceleration of the universe, 
so during inflation the bound 
$\epsilon_H \ll 1$ is fulfilled, 
and inflation ends once $\epsilon_H=1$.
Taking account that $\epsilon_H=q+1$, one obtain 
from (\ref{qcj}) that
\begin{equation}
\label{eps}
\epsilon_H =
\frac{3}{2}\frac{\text{dn}^{2}\left[(1+\alpha)\,\Phi\right]\,
\text{sn}^{2}\left[(1+\alpha)\,\Phi\right]}{\text{cn}^{2}\left[(1+\alpha)\,\Phi\right]}.
\end{equation}
The second Hubble hierarchy $\eta_{H}$
is defined as
\begin{equation}
\label{eta}
\eta_H\equiv -\frac{d\, \ln H_{,\phi}}{d\,\ln a}=\left(\frac{m_p^2}{4\pi}\right)\left(\frac{H_{,\phi \phi}}{H}\right),
\end{equation}
where plugging the Chaplygin--Jacobi scalar field becomes
\begin{equation}
\label{etaJ}
\eta_H=\epsilon_H\left\{ \frac{1+\alpha \,\textrm{cn}^2[(1+\alpha)\Phi]}{\textrm{sn}^2[(1+\alpha) \Phi]}+\frac{(1+\alpha)\,k'}{\textrm{dn}^2[(1+\alpha) \Phi]}\right\}.
\end{equation}
Finally, the third Hubble hierarchy parameter, $\xi_H^2$ is defined by
\begin{equation}
\label{xi}
\xi_H^2\equiv \left(\frac{m_p^2}{4\pi}\right)^2\,\left(\frac{H_{,\phi}\,H_{,\phi \phi \phi}}{H^2}\right),
\end{equation}
which becomes
\begin{equation}
\label{xiJ}
\xi_H^2=\epsilon_H^2\left\{
\frac{(2\alpha^2+7\alpha+6)\,k'}{\textrm{sn}^2\,[(1+\alpha)\Phi]\,\textrm{dn}^2\,[(1+\alpha)\Phi]}+\frac{3}{2\epsilon_H}
\left[(2k-1)-k\,\alpha\,(1+2\alpha)\,\textrm{cn}^2\,[(1+\alpha)\Phi]\right]
\right\}.
\end{equation}
The quantum fluctuations produce a power spectrum of scalar density fluctuations
of the form \citep{Guth:1982ec,Bardeen:1983qw}
\begin{equation}
\label{sdf}
\mathcal{P}_{\mathcal{R}}(\widehat{ \rm{k}})=\left(\frac{H}{|\dot{\phi}|}\right)^2
\left(\frac{H}{2\pi}\right)^2 \bigg\vert_{aH=\widehat{ \rm{k}}}.
\end{equation}
This perturbation is evaluated when a given mode $\widehat{ \rm{k}}$
crosses outside the horizon during inflation, i. e.
at $a H=\widehat{ \rm{k}}$. These modes do not evolve
outside the horizon, so we can assume they keep 
fixed value after crossing the horizon during
inflation.
In order to obtain some comparison with 
the available observational data, we introduce
the scalar spectral index $n_s$ defined as
\begin{equation}
\label{ns}
n_s-1\equiv \frac{d \ln \mathcal{P}_{\mathcal{R}}}{d \ln \widehat{ \rm{k}}}.
\end{equation}
After a brief calculation, one obtains that
\begin{equation}
\label{ns}
n_s-1= -4 \,\epsilon_H+2\,\eta_H,
\end{equation}
so, using Eqs. (\ref{eps}) and (\ref{etaJ}), the spectral index becomes

\begin{equation}
\label{siJ}
n_s-1=3\left\{\frac{\alpha\,k'}{\textrm{cn}^2\,[(1+\alpha)\Phi]}
-(2k-1)+(2+\alpha)\,k\,\textrm{cn}^2\,[(1+\alpha)\Phi]\right\}.
\end{equation}
As a first observation, 
note that the above equation 
allows to write $\textrm{cn}^2\,[(1+\alpha)\Phi]$ 
as a function of $n_s$. 
By solving the quadratic equation on
$\textrm{cn}^2\,[(1+\alpha)\Phi]$ 
is straightforward to find that
\begin{equation}
\label{cnns}
\zeta(n_s, \alpha, k)\equiv \textrm{cn}^2\,[(1+\alpha)\Phi]=\frac{\frac{1}{3}(n_s-1)+(2k-1)}{2\,k\,(2+\alpha)}
\left\{1\pm\sqrt{1-\frac{4\,k\,k'\,\alpha\,(2+\alpha)}{{\left[\frac{1}{3}(n_s-1)+(2k-1)\right]^2}}}\right\}.
\end{equation}
From here, it is no hard to see that 
$n_s$ get the value equal 
to one when the dimensionless 
scalar field get the value
\begin{equation}
\label{fins1}
\Phi_{n_s=1}=\frac{F\left[\arcsin\left(\pm \sqrt{\chi_{\pm}}\right),\,k\right]}{1+\alpha},
\end{equation}
where
\begin{equation}
\label{chi}
\chi_{\pm}=\frac{2k-1}{2\,k\,(2+\alpha)}
\left\{1\pm\sqrt{1-\frac{4\,k\,k'\,\alpha\,(2+\alpha)}{{(2k-1)^2}}}\right\}.
\end{equation}

\begin{figure}
  \begin{center}
    \includegraphics[width=85mm]{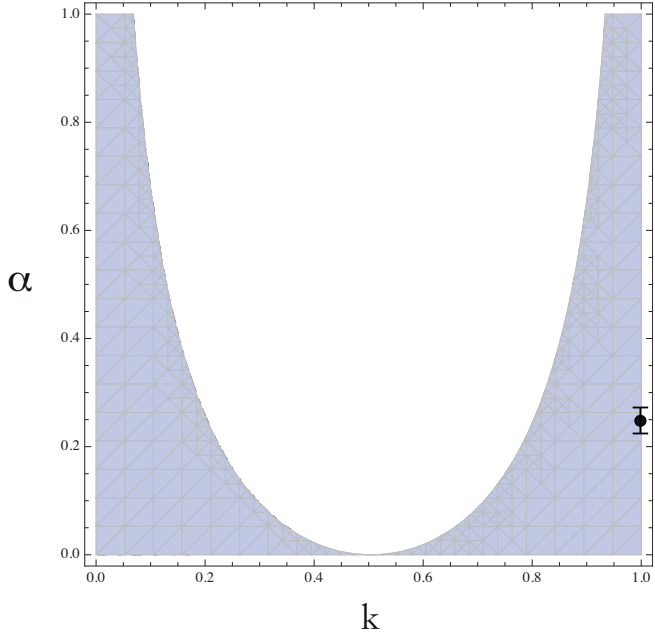}
  \end{center}
  \caption{Plot of the transcendental relation
  between the $\alpha$ parameter and the modulus $k$
 for the generalized Chaplygin--Jacobi gas (\ref{nrns}). 
 The shaded region corresponds to pairs 
 that are permitted by the observational data, 
 including the value found by del Campo using $k=1$, $\alpha
 =0.2587\pm 0.0009$ \cite{delcampo13}. 
 In this graph 
has been used $n_{run}=-0.003\pm0.007$ and $n_s=0.968\pm0.006$
from the {\it Planck} collaboration \cite{planck15}.}
    \label{planck15}
\end{figure}

On the other hand, one can introduce the 
{\it running scalar spectral index} defined by
$n_{run}\equiv \frac{d n_s}{d \ln \widehat{ \rm{k}}}$, 
which results to be
\begin{equation}
\label{rssi}
n_{run}=10 \,\epsilon_H\,\eta_H-8\,\epsilon_H^2-2\,\xi_H^2.
\end{equation}
Therefore, using Eqs. (\ref{eps}), (\ref{etaJ}) 
and (\ref{xiJ}) one get that
\begin{equation}
n_{run}=-6(1+\alpha)\left\{k'\,\alpha\, \textrm{nc}^2\,[(1+\alpha)\Phi]
-k\,(2+\alpha)\,\textrm{cn}^2\,[(1+\alpha)\Phi] \right\}\,\epsilon_H.\label{rssJ}
\end{equation}
Note that using eq. (\ref{cnns}) into eq. (\ref{rssJ})
one can obtain the running scalar spectral index as a function 
of scalar spectral index
\begin{equation}
\label{nrns}
n_{run}=-\frac{9(1+\alpha)(1-\zeta)(k'+k\,\zeta)[k'\,\alpha-k(2+\alpha)\zeta^2]}{\zeta^2},
\end{equation}
 and then, giving 
their value from the Planck collaboration \cite{planck15},
$n_{run}=-0.003\pm 0.007$ and $n_s=0.968\pm 0.006$,
it is possible to obtain a region of validity for the transcendental
relation between $\alpha$ and the modulus $k$
which is showed in FIG. \ref{planck15}.
Based in this results, 
the cases $k\approx 0$ and $k\approx 1$
are favoured in all range $0\leq \alpha \leq 1$.
The situation is completely different when
the modulus $k$ tends to $0.5$ where 
small values of the GCG parameter are favoured.

\begin{figure}
  \begin{center}
    \includegraphics[width=85mm]{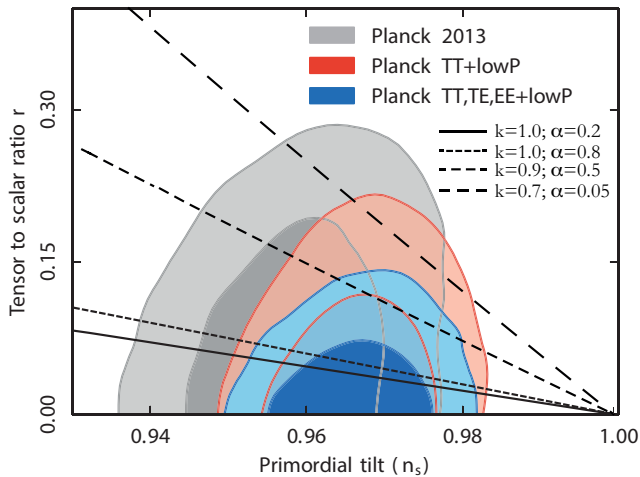}
  \end{center}
  \caption{Plot of the parameter $r$ 
  as a function of the scalar spectral index $n_s$
 for the generalized Chaplygin--Jacobi gas (\ref{rvsns}). 
 The marginalized joint $68\%$ and $95\%$ confidence level regions
 using Planck TT + low P, Planck TT, TE, EE + low P and 
 Planck 2013 data release with $\widehat{ \rm{k}}=
 0.05h$ Mpc$^{-1}$ are shown  \cite{planck15}. 
To perform a consistent comparison, the values of $r$ have been normalized in such a way
that  $r=0$ when $n_s =1$, and $r=16\epsilon_H$ is used. Note that this
is four times the normalization used in this work, see equation (\ref{defr}).}
    \label{rns15}
\end{figure}

It is known that not only scalar curvature perturbations 
are generated during inflation. In addition quantum 
fluctuations generate transverse-traceless tensor 
perturbations \cite{Mukhanov:1990me}, which do not couple to matter. 
Therefore they are only determined by the dynamics 
of the background metric. The two independent 
polarizations evolve like minimally coupled 
massless fields with spectrum
\begin{equation}
\label{pt}
\mathcal{P}_{\mathcal{T}}=\frac{16\pi}{m_p^2}
\left(\frac{H}{2\pi}\right)^2 \bigg\vert_{aH=\widehat{ \rm{k}}}.
\end{equation}
In the same way as the scalar perturbations, it is 
possible to introduce the gravitational wave spectral index
$n_T$ defined by $n_T\equiv \frac{d\ln \mathcal{P}_{\mathcal{T}}}{d\ln \widehat{ \rm{k}}}$, 
that becomes $n_T =-2\epsilon_H$ . 
At this point, it is possible to introduce the tensor--￼to--scalar 
amplitude ratio $r \equiv \frac{\mathcal{P}_{\mathcal{T}}}{\mathcal{P}_{\mathcal{R}}}$
which becomes
\begin{equation}
\label{defr}
r=4\epsilon_H,
\end{equation}
therefore one arrives to
\begin{equation}
\label{rcj}
r=6 \frac{\text{dn}^{2}\left[(1+\alpha)\,\Phi\right]\,
\text{sn}^{2}\left[(1+\alpha)\,\Phi\right]}{\text{cn}^{2}\left[(1+\alpha)\,\Phi\right]}.
\end{equation}
With the help of eq. (\ref{cnns}) is possible 
to find the relation between the tensor--￼to--scalar 
amplitude ratio and the scalar spectral index, which reads
\begin{equation}
\label{rvsns}
r(n_s, \alpha, k)=6\frac{(k'+k\,\zeta)(1-\zeta)}{\zeta},
\end{equation}
which is contracted with the measurement 
from {\it Planck collaboration} \cite{planck15} as is showed in 
Figure \ref{rns15}.

\section{Final remarks}
This paper presents an inflationary universe model 
in which the inflaton field is characterized
by an EoS corresponding to a generalized 
Chaplygin--Jacobi gas (GCJG), i.e. 
$p_{\phi}=-\frac{B\,k}{\rho_{\phi}^{\alpha}}
-2\,k'\,\rho_{\phi}+\frac{k'}{B}\,\rho_{\phi}^{2+\alpha}$, 
where $\alpha$
is the GCG parameter, 
and was considered to lie in the range $0 \leq \alpha \leq 1$,
and the $k$ parameter is the {\it modulus} of the elliptic function which lie 
in the range $0\leq k \leq 1$. 
In this study  the kinematical 
evolution was described 
using the Hubble parameter given by 
$H(\phi, k)=H_0\,\textrm{nc}^{\frac{1}{1+\alpha}}\left[(1+\alpha)\,\Phi\right]$.
From here, all relevant quantities that describe inflation
(scale factor, number of e-folds, deceleration parameter, etc.)
were obtained in terms of Jacobian elliptic functions, allowing
express them in an analytical form. 

As a direct application of these analytical results, 
it is possible to obtain an expression relating
the running scalar spectral index 
with the scalar spectral index (cf.  equation (\ref{nrns})). 
Using {\it Planck} 2015 data \cite{planck15},
$n_{run}=-0.003\pm 0.007$ and $n_s=0.968\pm 0.006$,
the transcendental relation
in the $\alpha$--$k$ space, showed in Figure \ref{rns15},
was obtained.
Notice del Campo \cite{delcampo13} 
obtained just one value  
$\alpha =0.2587\pm 0.0009 $ using the {\it Planck} 2013
data \cite{Planck:2013jfk}, which is in agreement
with the present results.
In the same way, a relation between
the tensor--to--scalar ratio parameter, $r$,
and $n_s$ is found (cf. equation (\ref{rvsns})),
and  contrasted with the marginalized joint 
68\% and 95\% confidence level regions using 
Planck TT+low P, Planck TT, TE, EE + low P and Planck 2013
release with $\widehat{ \rm{k}}=
 0.05h$ Mpc$^{-1}$ \cite{planck15}.
Based on these comparisons, it can be argued that 
the GCJG model is very auspicious regarding the available
observational tests.

Finally, since this generalization introduces a new parameter (the modulus $k$), 
it provides a viable opportunity to improve 
confrontations with the observational data.
Thus, this model offers new opportunities
to further investigate on the accelerated
stage and interacting models.

\appendix
\section{ A brief review of Jacobian elliptic functions}\label{app:jef}
As starting point, let us consider
the elliptic integral \cite{byrd,hancock,Armitage,Mey0l,tablas}
\begin{equation}\label{a1}
u(y, k)\equiv u=\int_0^y\frac{dt}{\sqrt{(1-t^2)(1-k\,t^2)}}\\
=
\int_0^{\varphi}\frac{d\theta}{\sqrt{1-k\sin^2\theta}}=
F(\varphi, k),
\end{equation}
where $F(\varphi, k)$ is the {\it normal elliptic integral of the first kind},
and $k$ is the {\it modulus}.
The problem of the inversion of this integral was
studied and solved by Abel and Jacobi, and leads
to the inverse function defined by
$y=\sin\varphi=\textrm{sn}(u, k)$
with $\varphi=\textrm{am}\, u$, and are called
{\it Jacobi elliptic sine} $u$ and {\it amplitude} $u$.

The function sn $u$ is an odd elliptic function of order two. It possesses
a simple pole of residue $1/k$ at every point congruent to
$i K'$ (mod $4K$, 2$i K'$) and a simple pole of residue
$-1/k$ at points congruent to $2K+i K'$ (mod 4 $K$, $2 i K'$),
where $K\equiv K(k)=F(\pi/2, k)$ is the {\it complete elliptic integral of the first kind},
$K'=F(\pi/2, k')$, and $k'=1-k$ is the {\it complementary modulus}.

Two other functions can then defined by $\textrm{cn}(u, k)=\sqrt{1-y^2}=\cos \varphi$,
which is called the {\it Jacobi elliptic cosine} $u$, and is an even function of order two;
$\textrm{dn}(u, k)=\sqrt{1-k\,y^2}=\Delta \varphi=\sqrt{1-k\,\sin \varphi}$, called
the {\it Jacobi elliptic delta} $u$ which is an even function.
The set of functions $\{\textrm{sn}\, u,\, \textrm{cn}\, u,\, \textrm{dn}\, u\}$
are called {\it Jacobian elliptic functions}, and 
take the following special values
\begin{eqnarray}
&&\textrm{sn} (u| 0)=\sin u,\quad \textrm{sn} (u| 1)=\tanh u,\\
&&\textrm{cn} (u| 0)=\cos u,  \quad \textrm{cn} (u| 1)=\textrm{sech}\, u,\\
&& \textrm{dn} (u| 0)=1,\, \qquad \,\,\textrm{dn} (u| 1)=\textrm{sech}\, u.
\end{eqnarray}
The quotients and reciprocal of $\{\textrm{sn}\, u,\, \textrm{cn}\, u,\, \textrm{dn}\, u\}$
are designated in {\it Glaisher's notation} by
\begin{eqnarray}
&&\textrm{ns}\, u=\frac{1}{\textrm{sn} \, u},\quad \textrm{cs}\, u=\frac{\textrm{cn} \, u}{\textrm{sn} \, u},\quad \textrm{ds}\, u=\frac{\textrm{dn} \, u}{\textrm{sn} \, u},\\
&&\textrm{nc}\, u=\frac{1}{\textrm{cn} \, u},\quad \textrm{sc}\, u=\frac{\textrm{sn} \, u}{\textrm{cn} \, u},\quad \textrm{dc}\, u=\frac{\textrm{dn} \, u}{\textrm{cn} \, u},\\
&& \textrm{nd}\, u=\frac{1}{\textrm{dn} \, u},\quad \textrm{sd}\, u=\frac{\textrm{sn} \, u}{\textrm{dn} \, u},\quad \textrm{cd}\, u=\frac{\textrm{cn} \, u}{\textrm{dn} \, u}.
\end{eqnarray}
Therefore, in all, we have twelve Jacobian elliptic functions.
Finally, some useful fundamental relations between Jacobian elliptic functions
are 
\begin{eqnarray}
&&\textrm{sn}^2 u +  \textrm{cn}^2 u=1,\\
&&\textrm{dn}^2 u+ k\, \textrm{sn}^2 u  =1,\\
&& \textrm{dn}^2 u-k\,\textrm{cn}^2 u=k',\\
&& \textrm{cn}^2 u+k'\,\textrm{sn}^2 u =\textrm{dn}^2 u.
\end{eqnarray}

\begin{acknowledgments}
This paper is dedicated to Sergio del Campo Araya (RIP). 
The author is very thank full to V\'ictor C\'ardenas, Ver\'onica Motta,
Osvaldo Herrera, Ram\'on Herrera and Winfried Zimdahl  
for their valuable comments which has improved the work.
This research is support  by Comisi\'on Nacional de Investigaci\'on Cient\'ifica 
y Tecnol\'ogica through FONDECYT grants No  11130695.
\end{acknowledgments}

\end{document}